\documentclass[acmsmall,screen=false,review=false,anonymous=false]{acmart}
\usepackage[utf8]{inputenc}

\AtBeginDocument{%
  \providecommand\BibTeX{{%
    \normalfont B\kern-0.5em{\scshape i\kern-0.25em b}\kern-0.8em\TeX}}}
    
\usepackage{caption}
\usepackage{subcaption}
\usepackage{breqn}

\usepackage{array}
\usepackage{multirow}

\newcommand\MyBox[2]{
  \fbox{\lower0.75cm
    \vbox to 1.7cm{\vfil
      \hbox to 1.7cm{\hfil\parbox{1.4cm}{#1\\#2}\hfil}
      \vfil}%
  }%
}

\setcopyright{acmlicensed}
\acmJournal{PACMHCI}
\acmYear{2022} \acmVolume{6} \acmNumber{CSCW2} \acmArticle{381} \acmMonth{11} \acmPrice{}\acmDOI{10.1145/3555106}

\usepackage{etoolbox}
\makeatletter
\pretocmd{\NAT@citexnum}{\@ifnum{\NAT@ctype>\z@}{\let\NAT@hyper@\relax}{}}{}{}
\makeatother

\def\citepos#1{\citeauthor{#1}'s \citep{#1}}

\begin{document}

\title[Many Destinations, Many Pathways]{Many Destinations, Many Pathways: A Quantitative Analysis of Legitimate Peripheral Participation in Scratch}

\author{Ruijia Cheng}
\email{rcheng6@uw.edu}
\orcid{0000-0002-2377-9550}
\affiliation{%
  \institution{University of Washington}
  \city{Seattle}
  \state{Washington}
  \country{USA}
  \postcode{98195}
}

\author{Benjamin Mako Hill}
\orcid{0000-0001-8588-7429}
\affiliation{%
  \institution{University of Washington}
  \city{Seattle}
  \state{Washington}
  \country{USA}
  \postcode{98195}}
\email{makohill@uw.edu}
\renewcommand{\shortauthors}{Ruijia Cheng \& Benjamin Mako Hill}

\received{July 2021}
\received[revised]{November 2021}
\received[accepted]{April 2022}

\begin{abstract}
Although informal online learning communities have proliferated over the last two decades, a fundamental question remains: What are the users of these communities expected to learn? 
Guided by the work of Etienne Wenger on communities of practice, we identify three distinct types of learning goals common to online informal learning communities: the development of domain skills, the development of identity as a community member, and the development of community-specific values and practices. 
Given these goals, what is the best way to support learning?
Drawing from previous research in social computing, we ask how different types of legitimate peripheral participation by newcomers---contribution to core tasks, engagement with practice proxies, social bonding, and feedback exchange---may be associated with these three learning goals. 
Using data from the Scratch online community, we conduct a quantitative analysis to explore these questions.
Our study contributes both theoretical insights and empirical evidence on how different types of learning occur in informal online environments. 
\end{abstract}



\maketitle


\section{Introduction}


In the last two decades, we have witnessed the proliferation of online communities that seek to promote informal learning through unstructured activities and interactions with others. For example, social computing scholars have documented the way users work together to learn creative and technical animation skills in communities such as \textit{NewGrounds} \cite{luther2009predicting}, writing and web development skills in fan communities such as \textit{FanFiction.net} and \textit{Archive of Our Own (AO3)} \cite{campbell2016thousands, fiesler2017growing}, and programming skills in creative coding communities such as \textit{Scratch} \cite{resnick2009scratch}. These communities do not offer fixed lesson plans and offer little in the way of formal instruction. They are interest-driven, open to all, and free to join. Members learn at their own pace while pursuing their personal passions.

Skeptics of informal learning have pointed out that many users of such systems spend a substantial portion of their time on these websites commenting and socializing. How effective can a programming community be in supporting learning if its users constantly chat with each other while spending relatively little time writing or reading code?
Proponents have responded that socialization promotes continued participation, a prerequisite to learning in informal settings where users are always at risk of not returning, and allows a collaborative process of learning and mentoring \citep{dasgupta2018wide, shorey2020hanging, campbell2016thousands}. Others have suggested that learning to socialize is an important learning outcome itself \cite{evans2017more, bruckman_2005, ito2013hanging}. An effective coding community could help its members develop important social skills in addition to technical ones \cite{brennan2011more}.

There remains a deep disagreement within the social computing community about how learning communities should be designed to maximize learning. We believe that resolving these conversations is difficult because researchers often do not fully answer the following questions: What do we expect members of informal learning communities to learn? What types of behavior should designers encourage and support for each learning goal? 

Our work begins to explore answers to these questions.
Guided by \citepos{lave1991situated} influential work on communities of practice (CoPs) and \citepos{wenger2002cultivating} more recent work on types of learning in CoPs, we describe three distinct categories of learning outcomes common to informal online learning communities: development of \textit{domain} skills, development of \textit{community} identity, and development of community \textit{practices}. Using these three outcomes, we explore how different types of legitimate peripheral participation (LPP) by newcomers---contribution to core tasks, engagement with practice proxies, social bonding, and feedback exchange---are associated with these different learning outcomes using data from the Scratch online community \cite{hill_longitudinal_2017}. We find that different types of participation are associated with learning outcomes differently. For example, making original projects as a newcomer is a good predictor of learning new computational concepts in the long term, while it is negatively associated with the formation of community identity and development of community practices.  

Our paper makes several contributions. First, we believe that ours is the first study to provide a quantitative analysis of LPP and learning outcomes in a CoP. In doing so, our work suggests that users' early participation in an online community is associated with long-term learning outcomes---but the successful types of participation associated with different types of learning outcomes will differ.
Second, we make an empirical contribution by analyzing 3 years of longitudinal data from Scratch. We believe that our work provides a template for future work that seeks to quantitatively investigate CoP theory in social computing in the context of a computational learning community.
Finally, our work informs the design of new systems by pointing out potential ways to facilitate different newcomer participation patterns supporting a range of learning outcomes.

\section{Background}
\label{sec:background}

\subsection{Communities of practice}
\label{sec:background:cop}

Over the last three decades, Jean Lave and Etienne Wenger's ideas around CoP have been among the most important theories used by social computing scholars to understand how learning happens in communities. 
The term \textit{community of practice} has been widely used in social computing scholarship to describe groups of people learning from each other while working toward a common interest or goal \cite[e.g.,][]{shrestha2021remote, marlow2014rookie, kou2018understanding, gilbert2016learning}. First introduced by \citet{lave1991situated} in 1991, CoP theory is grounded in ethnographic observation of apprenticeship relationships in communities of Liberian tailors, Mayan midwives, US Navy quartermasters, nondrinking alcoholics, and US supermarket meat cutters.




Learning in CoPs is described as occurring through LPP, the phenomenon through which newcomers begin to participate in a group by helping out with tasks that are easy and low-risk but still valuable and important. For example, a new midwife may begin by boiling water and cleaning scissors for other midwives. Through observing more experienced members performing more complex and higher-stake tasks---and by practicing themselves in various ways---novices move from the periphery to more central roles.

\subsection{Applications of CoP theory in social computing research}

Although created to explain learning through traditional apprenticeships offline, the HCI and CSCW communities have embraced the CoP framework. Today, it is one of the most influential and highly cited theories related to learning in social computing. CoP has been widely adopted by social computing scholars because it is a good match for informal and individualized forms of learning that occur in many online settings.

Despite being cited hundreds of times in social computing scholarship---and tens of thousands of times in general---there have been very few attempts to explore or test CoP theory quantitatively. We are not aware of any such attempts in social computing scholarship.
Furthermore, the work of applying CoPs to online settings, where groups are often extremely porous and diffuse, has required new theoretical work and argumentation. For example, \citet{gruzd2011imagining} have argued that a single social media user can form a personal or ``imaginary'' CoP with other users in their network. Using a similar line of thinking, data scientists on Twitter who engage with the hashtag ``\#TidyTuesday'' have been theorized to constitute a CoP because they share context, common interests, and a process of collaborative knowledge advancement and because they ask questions and interact with more experienced users \citep{shrestha2021remote}.

A wide range of other examples of online CoPs that have been identified and studied in social computing includes design professionals on online critique platforms \citep{marlow2014rookie, kou2018understanding}, a Facebook group of Airbnb hosts who learn about new features and hosting strategies from each other \citep{holikatti2019learning}, and fan fiction authors who gather online to develop and maintain the fan fiction website Archive of Our Own (AO3) \cite{fiesler2017growing}. 
In these examples and others, online CoPs are generally described as occurring in \textit{affinity spaces} where members with similar interests and identities contribute to a shared collection of knowledge in a distributed manner \cite{gee2005semiotic}. An example of an online affinity space that is frequently described as a CoP is the Scratch online community---the empirical context of our study \cite{monroy-hernandez_scratchr:_2007}.

\subsection{Types of learning in CoPs}


One limitation in Lave and Wenger's initial account is that it is vague about what exactly is being learned in a given CoP. This is an important omission because CoP members typically learn a range of different things as they become more experienced. For example, becoming a Mayan midwife in the Yucatan Peninsula involves much more than learning technical midwifery skills. It also involves learning about the Mayan midwifery community and the specific norms and values that shape midwifery in the Yucatan.
In later work, \citet{wenger2002cultivating} attempted to address this omission by identifying three types of learning in CoPs: learning about \textit{domain}, \textit{community}, and \textit{practice}.
We present each type of learning in the following paragraphs. 


First, learning about a \textit{domain} refers to the acquisition of knowledge and skills necessary for a person to carry out the core tasks at the heart of a CoP. Many scholars of online communities are interested in how communities use LPP to support learning about some domain of knowledge and domain learning is often assumed to be the only learning goal in a CoP. For midwives, learning about a domain involves gaining knowledge related to successfully delivering infants. Depending on the community, domain knowledge may involve skills related to computer programming \cite{resnick2009scratch, Krogha2003CommunityJA}, fan fiction writing \cite{campbell2016thousands}, encyclopedia article editing \cite{halfaker2013making}, and so on. In the specific context of computational learning communities such as Scratch, domain learning means learning computational concepts and related programming skills \cite{resnick2009scratch}. 

The second type of learning involves the development of identity as a member of the \textit{community}. This involves developing relationships, affinities, and a sense of belonging.
As learners are accepted by older members of the community, they gradually form an identity as a member of the community.
The development of these relationships and affinities would play out similarly in various settings and typically involves the knowledge of other community members and the development of a sense of membership and commitment to the community \cite{mcmillan1986sense}. 

Third and finally, learning a \textit{practice} means assimilating ``cultural artifacts, norms, and values'' developed in the community over time \citep{barab2000practice}. By moving from peripheral to central forms of participation, learners develop by adjusting their social, work, and contribution style to match what is accepted and appreciated by community members. The specific reasons why one would be seen as accepted and appreciated in a CoP are very community-specific. However, indicators of practice development will typically involve expressions of appreciation and respect from others. For example, for midwives, it might involve the authority to lead or manage other midwives \cite{lave1991situated}. On Fanfiction.net, signs of practice development may involve high ratings and positive comments \cite{campbell2016thousands}. On Scratch, it might involve ``loves'' (i.e., likes) given by other users \cite{brennan2013imagining}.

\subsection{How does LPP promote different types of learning in CoPs?}
\label{sec:lpptypes}

Just as CoPs promote multiple types of learning, they also allow multiple types of participation. Although many studies of online activity reduce behavior to unidimensional concepts like ``engagement,'' there are many types of LPP that occur in CoPs. To identify different types of LPP relevant to online communities, we conducted a detailed search of the literature on CoPs in CSCW and social computing venues.
We did not find any previous attempt to enumerate or classify different types of LPP in social computing. However, through our reading of the literature, we were able to identify four distinct types of LPP that are frequently discussed: contribution to core tasks, engagement with practice proxies (an important type of activity in CoP theory, which we will explain in detail below), the formation of social bonds, and feedback exchange. We do not claim that these four types are strictly mutually exclusive; nor do we claim that they form a comprehensive list. We merely offer them as examples of four different types of participation that capture some of the diversity of LPP. We discuss the limitations of our work based on our necessarily arbitrary identification of the types of LPP in §\ref{sec:threats}.


The first type of LPP we identified, \textit{contribution to core tasks}, refers to the work of newcomers toward a community's explicit goal. Examples of contributions to core tasks include editing articles on Wikipedia, submitting code in an open source software project, and creating programming projects on Scratch. The second type of LPP is engagement with \textit{practice proxies}. The term practice proxies refers to activities that allow newcomers to observe and participate in the socially salient aspects of others' unfolding work practices \cite{mugar2014planet}. For example, in CoPs for online citizen science, new users can learn to contribute by engaging with project documentation left by others \cite{mugar2014planet}. The third type of LPP is \textit{social bonding}. In \citepos{lave1991situated} study of an alcoholics social group, the most important activities involved members forming interpersonal bonds and becoming friends. In online CoPs, social bonding is often enabled by social media features such as friending and following. The fourth form of LPP is participation in \textit{feedback exchange}. Feedback exchange in online CoPs often takes place through public commenting on artifacts shared by community members.

Because participants in online groups learn different types of things, it stands to reason that the most effective forms of LPP for the promotion of learning might vary depending on the type of learning. In other words, not all forms of peripheral participation will be equally likely to help a newcomer develop in terms of every desirable learning outcome. 
For example, peripheral participation in socializing activities might help newcomers build knowledge about the community, but might not contribute to their knowledge of domain skills.
Consequently, we might not expect that socializing in a coding-focused CoP would necessarily make one a better programmer. 
Similarly, contribution to the core task of programming may not help to build social knowledge about the community and its members. Indeed, given limited time and resources, support for one learning outcome might come at the expense of others. For example, a study of the Scratch community showed that socialization among members in the comment section of Scratch projects can sometimes drive discussion away from programming related topics \citep{shorey2020hanging}.

\begin{figure*}[t]
  \centering
  \includegraphics[width=1\linewidth]{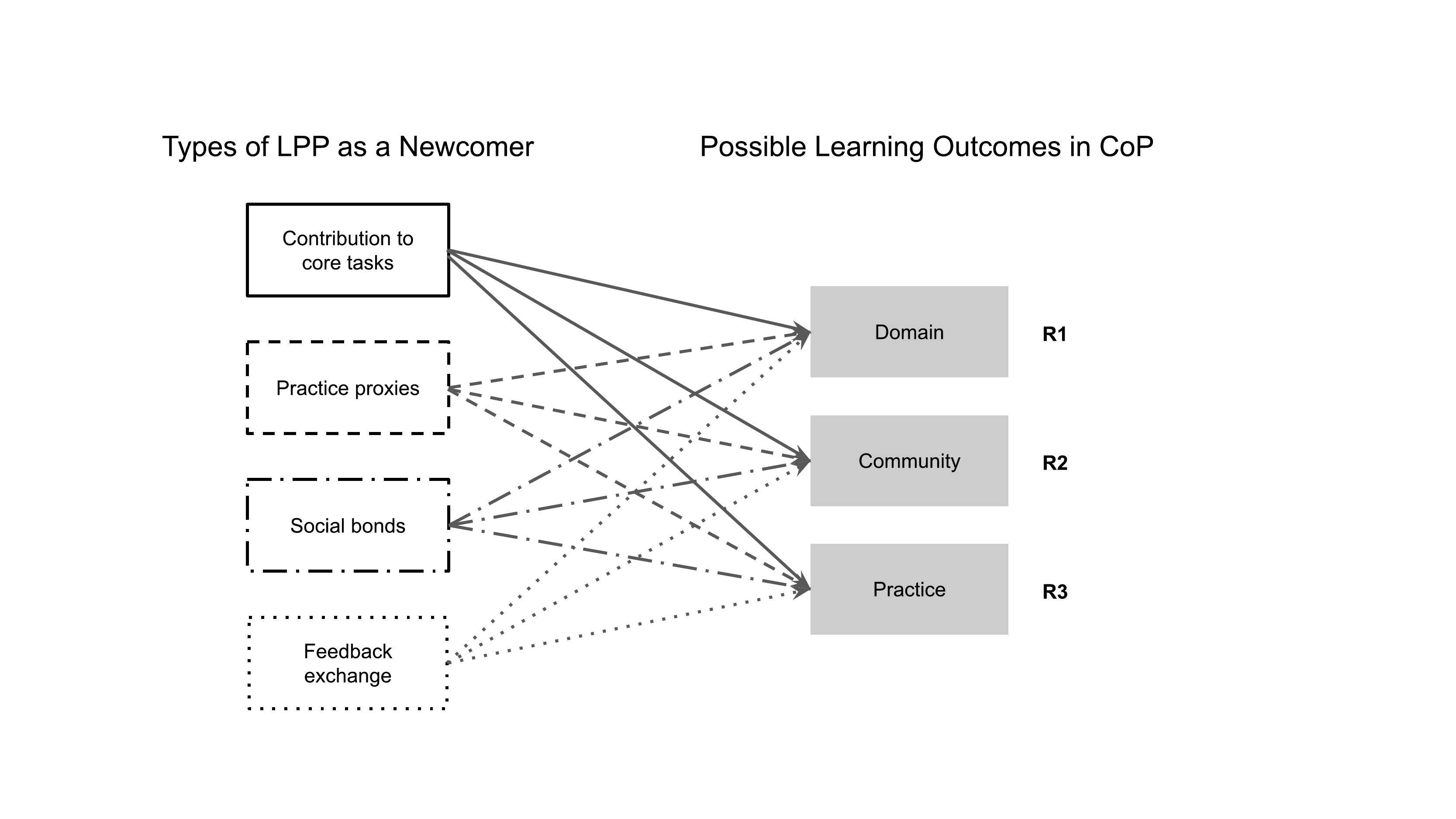}
  \caption{Our research examines which type of newcomers' LPP is associated with which learning outcome in CoP.}
  \Description{There are two columns in this figure. In the left column, on the top it says ``Types of LPP as a Newcomer.'' Underneath that are four boxes, from top to bottom: ``Contribution to core tasks,'' ``Practice proxies,'' ``Social bonds,'' ``Feedback exchange.'' In the right column, on the top it says ``Possible Learning Outcomes in CoP.'' Underneath that are three boxes, from top to bottom: ``Domain'' (R1) ``Community,'' (R2) ``Practice'' (R3). 12 arrows are drawn in total from each of the four left boxes to each of the three right boxes.}
  \label{fig:research questions}
\end{figure*}

Inspired by the original CoP literature that focused primarily on the experience of newcomers and their progression in the community \cite{lave1991situated}, we specifically focus on the participation of newcomers and their long-term learning outcomes. Figure \ref{fig:research questions} includes the four types of newcomer LPP we have described on the left and \citepos{wenger2002cultivating} three types of learning in CoPs on the right.
We draw 12 left-to-right arrows that represent all possible direct pathways between our four types of LPP and \citepos{wenger2002cultivating} three learning outcomes. In the following sections, we explain our research questions (R1 to R3) and related work on how different types of LPP contribute to the three learning outcomes. 

\subsubsection{Domain}
The social computing literature suggests that domain knowledge learning can be influenced by different types of LPP, although the specific effects remain unclear. For example, many studies suggest that by contributing to core tasks, newcomers can learn technical details and gain confidence. For example, new Wikipedia editors often start by making edits on topics that they are familiar with \cite{bryant2005becoming}. In the context of computational learning, novice programmers involved in open source projects often begin by reporting bugs and requesting features \cite{Krogha2003CommunityJA} and fanfiction authors learning to program contribute to the development of a fanfiction website by engineering very small features \cite{fiesler2017growing}. However, some also argue that contributing to core tasks can be stressful and overwhelming to newcomers, especially for those without much previous exposure to the domain. 

Alternatively, newcomers can learn domain knowledge and skills with the help of practice proxies. For example, social media functionality that allows community members to view and download artifacts shared by others has been described as a method by which practice proxies can build domain knowledge among new designers \cite{marlow2014rookie}. Similarly, new Wikipedia editors rely on practice proxies by reading existing pages and page histories before making their initial contributions to the encyclopedia as a way of learning where, when, and how to edit \cite{bryant2005becoming}. In the context of computational learning, the evidence of the relationship between engagement with practice proxies and domain learning is mixed. While research shows that Scratch users who create remixes of others' projects (a common practice proxy in computational learning) will learn more computational concepts as they continue to participate in the community \cite{dasgupta2016remixing}, they tend to display less innovation and originality in the programs they make \cite{hill2013cost}. 

Furthermore, while many computational learning platforms support socialization and feedback exchange, how these activities contribute to computational learning remains unclear. To explore how different types of LPP contribute to domain knowledge learning in CoPs, we ask the research question \textbf{R1:} How do different types of LPP affect domain knowledge learning? 

\subsubsection{Community}
Qualitative evidence has shown that \textit{engagement with practice proxies}, forming \textit{social bonds}, and participating in \textit{feedback exchange} can help newcomers develop an identity as a member of their \textit{community}.
Engagement with practice proxies can not only show newcomers how to master domain skills, but also provide newcomers with a sense of belonging. \citet{lave1991situated} describe how novice meat cutters in a US supermarket who lacked practice proxies had a low rate of interaction with more experienced butchers and did not feel like they were participating in a community to learn skills. In the context of Scratch, many users engage in remixing with the sole purpose of broadcasting each other's work and being part of a social group \cite{hill2013cost}. In the online CoP of R data scientists on Twitter, a large portion of newcomers join to build connections with others in the R community \cite{shrestha2021remote}.
Furthermore, social computing research has shown that newcomers with connections to others tend to remain active in the community longer \cite{kraut_building_2011}. 

Previous research has indicated that newcomers' engagement in feedback exchange can also help users develop as community members. Feedback exchange in online CoPs has been described as a form of distributed mentoring in which mentoring relationships are not bounded by time, geographic space, and differences in expertise \cite{campbell2016thousands}. Similarly, an experiment on reader engagement on Wikipedia has shown that newcomers showed higher long-term participation rates when asked to provide feedback on articles \cite{halfaker2013making}. 
To obtain quantitative evidence for how different types of LPP support members to develop community identity, we explore the research question \textbf{R2:} How do different types of LPP affect the development of community identity? 

\subsubsection{Practice}
Few papers in the social computing literature have directly studied how \textit{practice} in a community---i.e., being able to understand and reproduce what is valued by CoP members---can be supported. Some qualitative studies have implied that feedback exchange could help members of online CoPs learn community practices. For example, in forums dedicated to online dating, users offer each other comments and advice on self-presentation to understand how others use dating sites and what behaviors are considered appropriate \cite{masden2015understanding}. Similarly, new Airbnb hosts critique each other's drafts of hosting descriptions to recognize effective hosting styles \cite{holikatti2019learning} and young UX enthusiasts comment on others' posts to help each other learn how to make an appropriate networking message in the community \cite{kou2018understanding}. To further explore if and how different types of LPP can help learners learn community practices differently, we ask the research question \textbf{R3:} How do different types of LPP affect the learning of community-specific values and practices? 


\section{Empirical Setting: Scratch Community}
\label{sec:empirical}

\begin{figure*}[t]
     \centering
     \begin{subfigure}[b]{0.48\textwidth}
         \centering
         \includegraphics[width=\textwidth]{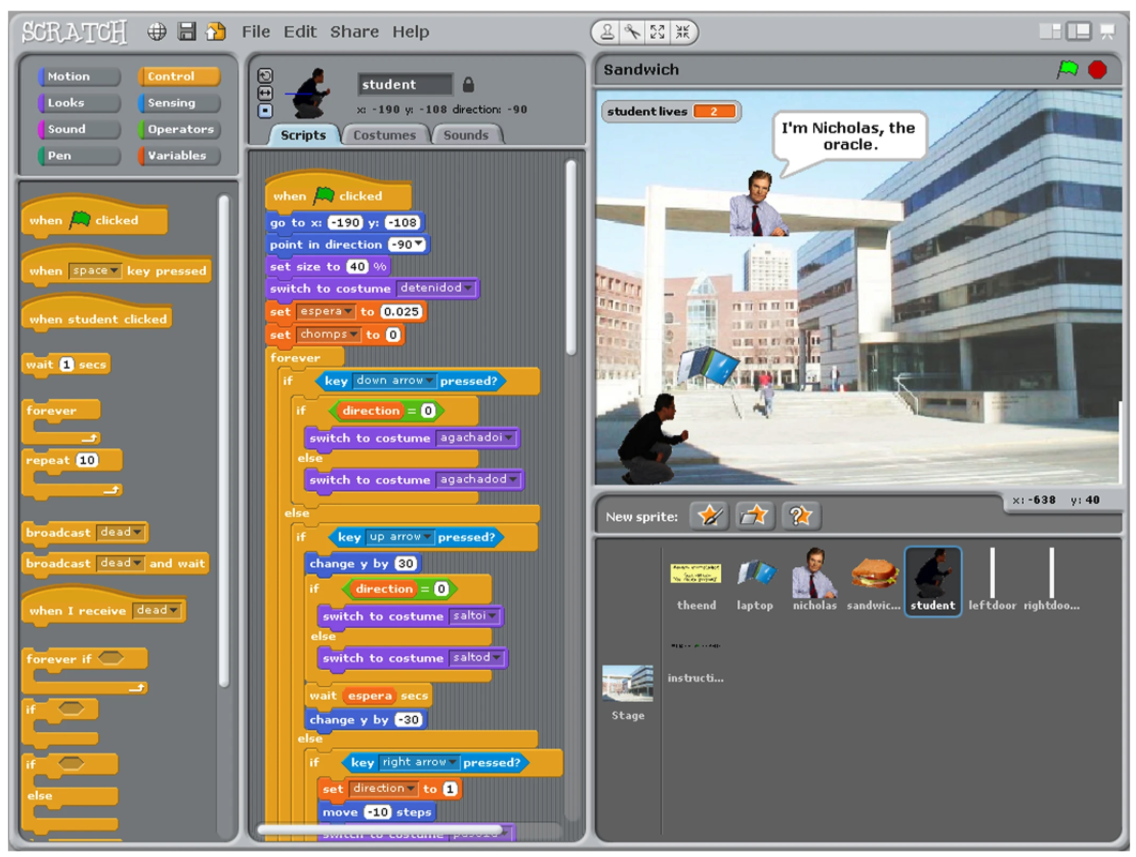}
         \caption{Example of Scratch code.}
         \label{fig:scratch_code}
     \end{subfigure}
     \hfill
     \begin{subfigure}[b]{0.48\textwidth}
         \centering
         \includegraphics[width=\textwidth]{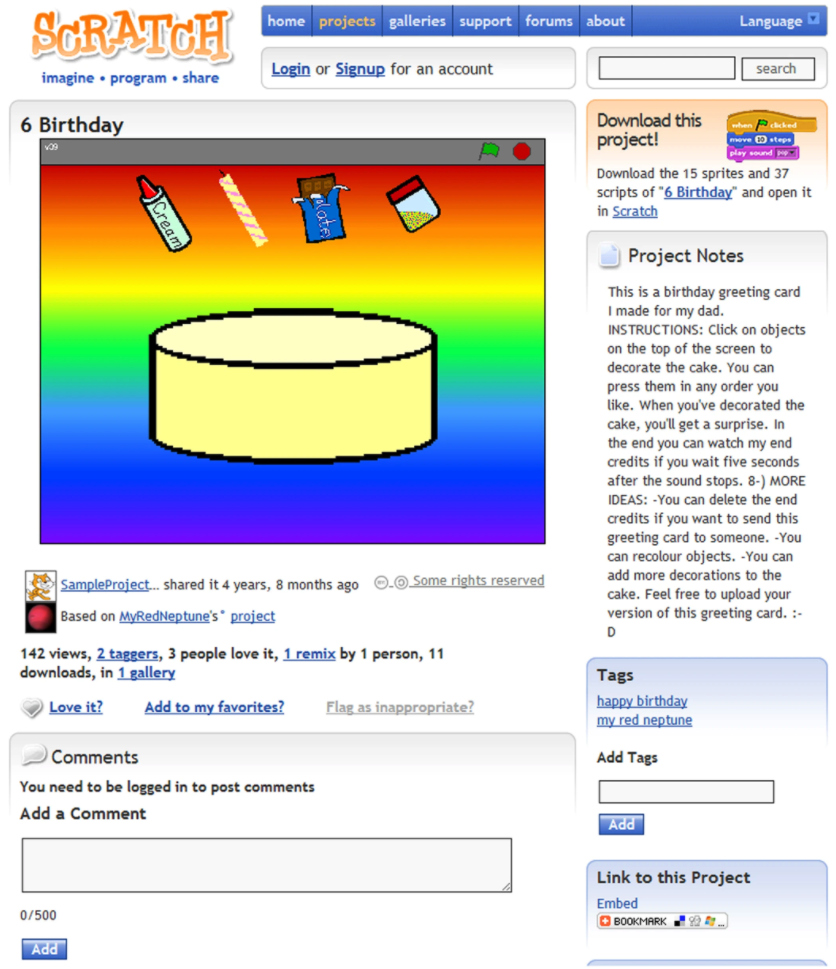}
         \caption{Social features and activities around the project.}
         \label{fig:scratch_social}
     \end{subfigure}
     \Description{Left figure: a screenshot of the Scratch programming interface. On the right is a stack of the visual programming blocks. On the right is the display of the project, where the background is a building and a character says ``I'm Nicholas, the oracle.'' Right figure: a screenshot of the webpage for a Scratch project. The project title is ``Birthday.'' The preview of the project includes a rainbow colored background, a cake, and four items above the cake: a tube that says ``cream,'' a candle, a chocolate bar, and a bag of sprinkles. Underneath the preview, it shows 142 views, 2 taggers, 3 people love it, 1 remix by 1 person, 11 downloads, in 1 gallery. Underneath those information is an empty comment box that says ``You need to be logged in to post comments'' and ``Add a Comment.''}
     \caption{Historical interfaces of Scratch and social features at the time when data used in our analysis were collected. Images obtained from \citet{hill_longitudinal_2017}.}
\end{figure*} 

We conducted our study using data from the Scratch community. Scratch was designed to support young people in learning to program using the Scratch programming language---a visual block-based language designed for children to learn basic programming \cite{resnick2009scratch, roque2012from}. Scratch was designed based on constructionist learning principles \cite{papert1980mindstorms, resnick1996pianos}. Programming primitives are represented by visual blocks that control the behavior of graphical objects on the screen called sprites. As shown in Figure \ref{fig:scratch_code}, users can drag and drop blocks together to build programs.

Although the Scratch programming language was designed for children to use on their own computers, the language has been integrated into a vibrant online community since 2007. Users in this community can view others' project and share their programming projects with others \cite{monroy-hernandez_scratchr:_2007}.\footnote{\url{https://scratch.mit.edu/}} As of July 2021, the Scratch online community has over 73 million registered users and over 79 million shared projects that span a diverse range of genres and themes. The large majority of Scratch users are between the ages of 8 and 16 years and the average age for new contributors is approximately 12 years.\footnote{All statistics about Scratch community activity and users are taken from the public information at: \url{https://scratch.mit.edu/statistics/}}

We chose Scratch for our study because it has been the center of several influential studies on informal learning in social computing and has been described as a CoP or site for situated learning in a range of previous studies \cite{dasgupta2016remixing, hautea2017youth, dasgupta2018wide}.
Beyond its importance in social computing scholarship, Scratch is one of the largest online communities for children learning to code.
Critically for our analysis, Scratch supports a variety of learning outcomes and forms of participation that allow us to test all three of our research questions and test the hypotheses associated with each arrow in Figure \ref{fig:research questions}.
Scratch users have access to practice proxies in the form of affordances that support users in viewing, editing, and building on others' code by downloading and remixing each other's projects \cite{monroy-hernandez_scratchr:_2007}. Previous studies have found evidence that remixing others' projects is associated with measures of learning about programming \cite{dasgupta2016remixing}. 
Scratch also supports users to build social connections by ``friending'' other users to form social networks and a shared sense of community \cite{brennan2011more, brennan2013imagining}. Scratch users can participate in feedback exchange by commenting on each others' projects, as shown in Figure \ref{fig:scratch_social}. For example, previous research has revealed collaborative debugging activities in project comments \cite{shorey2020hanging}. Finally, users can demonstrate social support by ``loving'' (upvoting) and ``favoriting'' (bookmarking) projects of others. We describe how these features map to our measure in the following section.


\section{Data and Measures}
\label{sec:data}


Our data are drawn from publicly available \textit{Scratch Research Dataset} (SRD), which includes comprehensive public data from the first 5 years of activity in Scratch between 2007 and 2012 \citep{hill_longitudinal_2017}. 
We further restrict our analysis to projects created between 2 July 2009 and 10 April 2012 because both programming and social affordances in the site were consistent during this period.\footnote{\url{https://en.scratch-wiki.info/wiki/Scratch_1.4}} Our analysis relies on the tables \textit{projects}, \textit{users}, \textit{pcomments}, \textit{gcomments}, \textit{project\_blocks}, \textit{favorites}, \textit{lovers}, \textit{viewers}, and \textit{friends} from the SRD.
Although this dataset is almost 10 years old, limiting ourselves to data from this period means that we can be more sure that the data are generated consistently and that others can reproduce our analysis using public data. To further support reproducibility, we have made the complete analytical code used in our analysis available.\footnote{https://dataverse.harvard.edu/dataset.xhtml?persistentId=doi:10.7910/DVN/0VXEMB} 
As in many other large online communities, Scratch users frequently register and participate only briefly before becoming inactive \cite{CROWSTON201889}.
Because we are interested in how users' activities as newcomers predict long-term outcomes, we created a user-level dataset with 121,149 users who were active for at least one day after their initial registration during our period of observation. 

For all of our research questions, we attempt to measure activity during two periods: users' time spent as newcomers and their time as established users. Following a series of previous studies on Scratch, we define the ``newcomer period'' of Scratch users as consisting of the first 14 days after each user creates their account on Scratch. Like participants in most online communities, Scratch users' transition from newcomers to established members is fuzzy and variable. From our own participation in the Scratch community, we sense that nearly all users would be considered newcomers on their first several days, but most users would no longer be considered newcomers after more than a month of activity.
We choose 14 days because it falls between this range and because it is embedded in one of Scratch's key platform policies in that ``Scratcher'' status is only granted to users who have been on the platform for at least 14 days.\footnote{\url{https://en.scratch-wiki.info/wiki/Scratcher}}
Within Scratch, users with Scratcher status are typically not considered newcomers by other community members and have access to more advanced programming features, such as the Scratch Cloud Variables \cite{dasgupta2018wide}. Other social computing studies have also defined the first 14-day period as the newcomer period and use user behaviors in this period to predict subsequent activities \cite{burke2009feed}. Because this choice of 14 days is necessarily arbitrary, we also repeated our analyses with 2, 7, and 30 days as the length of the newcomer period in a series of robustness checks. All these analyses lead to similar results and conclusions. We include Table \ref{table:distribution} to give a sense of the distribution of activity in the first 14 days in our sample. A pattern---common in online community data from many sources---is that our measures are heavily right-skewed. Although some users are very active, the median number for most measures is 0.

\begin{table}[h]
\centering 
\begin{tabular}{lcccc} 
\hline
Measure & \multicolumn{1}{c}{Mean} & \multicolumn{1}{c}{St. Dev.} & \multicolumn{1}{c}{Median} & \multicolumn{1}{c}{Range} \\ 
\hline 
Original projects made in the first 14 days & 2.200 & 4.515 & 1 & [0, 222] \\ 
Remixes made in the first 14 days & 0.656 & 2.219 & 0 & [0, 130] \\ 
Friends made in the first 14 days & 1.269 & 9.115 & 0 & [0, 1,487] \\ 
Comments made in the first 14 days & 4.516 & 26.640 & 0 & [0, 2,450] \\ 
Friends received in the first 14 days & 0.832 & 2.877 & 0 & [0, 206] \\ 
Loves received in the first 14 days & 0.492 & 2.668 & 0 & [0, 261] \\ 
Views received in the first 14 days & 11.133 & 34.935 & 3 & [0, 4, 161] \\ 
Remixes received in the first 14 days & 0.111 & 1.528 & 0 & [0, 446] \\ 
Comments received in the first 14 days & 2.706 & 16.079 & 0 & [0, 1,418] \\ 
Favorites received in the first 14 days & 0.271 & 1.529 & 0 & [0, 160] \\ 
CT concepts displayed in the first 14 days & 2.886 & 2.364 & 3 & [0, 6] \\ 
Total active duration (in days) & 110.656 & 171.079 & 36 & [1, 1,002] \\ 
Total new CT concepts & 4.281 & 1.981 & 5 & [0, 6] \\ 
\hline \\ 
\end{tabular} 
\caption{Descriptive statistics for a range of measures of user activity in Scratch including all activities that factor into our analysis.} 
  \label{table:distribution} 
\end{table} 

\begin{table}[t]
\centering 
\begin{tabular}{lccccc} 
\hline
Variable & \multicolumn{1}{c}{Mean} & \multicolumn{1}{c}{St. Dev.} & \multicolumn{1}{c}{Median} & \multicolumn{1}{c}{Range}  & Type\\ 
\hline 
\\
Independent variables: \\
\\
Made\_original\_project? & 0.710 & - & 1 & \{0,1\} & binary \\ 
Made\_remix? & 0.271 & - & 0 & \{0,1\} & binary\\ 
Made\_friend? & 0.247 & - & 0 & \{0,1\} & binary\\ 
Made\_comment? & 0.344 & - & 0 & \{0,1\} & binary\\
\\
Control variables: \\
\\
Has\_been\_friended? & 0.271 & - & 0 & \{0,1\} & binary\\ 
Has\_been\_loved? & 0.170 & - & 0 & \{0,1\} & binary\\ 
Has\_been\_viewed? & 0.725 & - & 1 & \{0,1\} & binary\\ 
Has\_been\_remixed? & 0.057 & - & 0 & \{0,1\} & binary\\ 
Has\_been\_commented? & 0.329 & - & 0 & \{0,1\} & binary\\ 
Has\_been\_favorited? & 0.116 & - & 0 & \{0,1\} & binary\\ 
CT\_concepts & 2.886 & 2.364 & 3 & [0,6] & count\\
\\
Outcome variables: \\
\\
Stayed? & 0.678 & - & 1 & \{0,1\} & binary\\ 
New\_CT\_concepts? & 0.396 & - & 0 & \{0,1\} & binary\\
Received\_new\_loves? & 0.044 & - & 0 & \{0,1\} & binary\\
\hline  \\
\end{tabular}
\caption{Distribution of variables used for regressions.} 
 \label{table:distribution_variables} 
\end{table} 

Because most of the variation in the variables is between 0 and 1, we transform most of these variables into simple dichotomous measures for analysis. Each of the measures used in our regression analysis is shown in Table \ref{table:distribution_variables}. 
Three variables---\textit{Stayed?}, \textit{New_CT_concepts?}, and \textit{Received_new_loves?}---are our binary outcome variables. The rest are independent variables and controls. 
All of these are drawn directly from the SRD or computed through merging and aggregating user-level activities across tables in ways that are self-explanatory and well documented in our code.

The most important exception to this is our measures related to computational thinking (CT) concepts. CT is defined as ``the thought processes involved in the formulation of problems and their solutions so that the solutions are represented in a form that can be carried out effectively by an information processing agent'' \citep{wing2008computational}. Referring to \citepos{brennan2012new} interpretation of CT and categorization of CT concepts, Scratch's designers designed programming blocks to embody specific CT concepts, including \textit{loops}, \textit{parallelism}, \textit{events}, \textit{conditionals}, \textit{operations}, and \textit{data} \cite{resnick2009scratch}. Our measure of \textit{CT concepts} captures the number of each distinct concept that a user displays in their projects made in their first 14 days. The specific CT concept-to-block mapping that we use is drawn from the study by \citet{dasgupta2016remixing} and can be found in our appendix in Table \ref{table:CT}.


\subsection{Measures}

\subsubsection{Dependent Variables}
Because Scratch does not make any systematic attempt to measure learning, an analysis like ours must rely heavily on proxy measures of our outcomes.
We constructed three outcome variables that correspond to our three research questions.
The outcome variable for \textbf{R1} measures the \textit{domain} being learned in Scratch---the computational skills. To construct a measure to capture the learning of computational skills, we follow the approach of previous quantitative studies on computational learning in Scratch. Our measure captures the size of the learners' repertoire in terms of the number of types of CT concepts a user has demonstrated in their projects \cite{scaffidi2012skill, yang2015uncovering, dasgupta2016remixing, dasgupta2018wide}. We construct an outcome measure to capture the learning of computational skills: \textit{New_CT_concepts?}, a binary variable of whether the user uses any new CT concepts in their projects after the first 14 days. 

For the outcome variable for \textbf{R2}, our proxy measure for \textit{community} membership is the duration of a user staying active in the community. Although this is not a direct measure of learning, CoP theory suggests that users who are more integrated into a community will stay longer.
We measure this outcome with a binary variable, \textit{Stayed?}, that captures whether users will participate in any recorded community activities after 14 days. In this case, activity can include sharing original or remixed projects, posting comments, favoriting, and/or friending. 

For \textbf{R3}, we want to test how participation in feedback exchange contributes to learning community-specific \textit{practices} and values. Because it is difficult to directly measure users' learning of community values, we explore H3 using a proxy that seeks to measure the Scratch community's positive reaction to projects made by a user.
Following previous work by \citet{hill2013cost}, we measure the reaction of the Scratch community to a user's projects as the number of loves received by that user.
Specifically, we construct the measure for the outcome variable in R3 as \textit{Received_new_love?}, a binary variable of whether the user received any loves on projects created after their first 14 days. 

\subsubsection{Independent Variables}
We construct four independent variables to measure the four types of LPP that we describe in §\ref{sec:lpptypes}.
First, 
we construct a measure for the LPP of contribution to core tasks. Since the Scratch community is designed to help children learn computer programming, the core task in Scratch is to write code and share original projects \citep{resnick2009scratch}. Therefore, we construct a binary measure, \textit{Made_original_project?}, that captures whether a user has shared an original project in their first 14 days. 
Second, we construct a measure for engagement with practice proxies. Drawing from previous literature on LPP in Scratch, we choose to operationalize practice proxies as \textit{Made_remix?}, a binary variable of whether the user remixed in their first 14 days.
Third, we construct a measure for social bonding. The most explicit way to form social bonds in Scratch is by adding other users as ``friends.'' In Scratch, friending means that a user has followed another user and will receive notifications when they post new projects.\footnote{\url{https://en.scratch-wiki.info/wiki/Friend}} Because we want to measure whether a user is actively participating in forming social bonds, we construct a binary variable, \textit{Made_friend?}, which captures whether a user has friended at least one other user in their first 14 days. 
Fourth, we constructed a measure for newcomers' participation in feedback exchange. In Scratch, feedback is exchanged mainly in the form of comments \cite{shorey2020hanging}. There are two types of comments in Scratch: comments on projects and on ``galleries.''
We sum these two types of comments into the total number of comments posted and received by each user. Because we want to ensure that we are measuring whether a user is actively engaged in feedback exchange, we construct a binary variable, \textit{Made\_comment?}, which captures whether a user has posted a comment in their first 14 days.

\subsection{Control variables}

Based on previous work, we add a series of control variables that capture reasons that users might differ in their learning outcomes beyond differences in the four types of LPP we identify.
For example, because previous literature has suggested that social support can affect newcomers' learning \cite{burke2009feed, kraut_building_2011, ford2016paradise}, we control for incoming social approval on projects, constructed as four binary variables---\textit{Has_been_loved?}, \textit{Has_been_favorited?}, \textit{Has_been_viewed?}, and \textit{Has_been_remixed?}---which measure whether a user received loves, favorites, views, and remixes on their projects, respectively, in their first 14 days on Scratch. We also construct two binary variables, \textit{Has_been_commented?} and \textit{Has_been_friended?}, to control for the effect of passively received feedback and social bonds. Because the programming experience of users before joining the Scratch community may affect users' learning of CT concepts and the reception of their projects within Scratch, we also included a control variable, \textit{CT_concepts} that measures the total number of unique CT concepts used by a user in their first 14 days on Scratch. This control is only relevant in R1 and R3.

\section{Analytic Plan}
\label{sec:analytic}

To answer R1 using the binary outcome variable \textit{New_CT_concepts?}, we fit a logistic regression model on the dataset of 121,149 users using the GLM function in R.\footnote{\url{https://www.rdocumentation.org/packages/stats/versions/3.6.2/topics/glm}} Because the distribution of the count control variable \textit{CT_concepts} is right-skewed (i.e., most users used 0 CT concepts), we use a started log transformation in all the models involving this control (i.e., $\log(x+1)$). Because our outcomes and all other measures are binary, our model is nonparametric, except for our \textit{CT\_concepts} variable. Our formal model for (M1) is as follows:

\begin{dmath}
\label{eq:m1b}
\log{\left(\frac{\hat{p}(\mathit{New\_CT\_concepts?})}{1-\hat{p}(\mathit{New\_CT\_concepts?})}\right)} = \beta_0 +\beta_1 \mathit{Made\_original\_project?} + \beta_2 \mathit{Made\_remix?} + \beta_3 \mathit{Made\_friend?} + \beta_4 \mathit{Made\_comment?}
+ \beta_5 \mathit{Has\_been\_friended?} + \beta_6 \mathit{Has\_been\_loved?} + \beta_7 \mathit{Has\_been\_viewed?} + \beta_8 \mathit{Has\_been\_remixed?} + \beta_9 \mathit{Has\_been\_commented?} + \beta_{10} \mathit{Has\_been\_favorited?} + \beta_{11} \log{(\mathit{CT\_concepts}+1)}
\end{dmath}


%

To answer R2 with the binary outcome variable \textit{Stayed?}, we fit a logistic regression model (M2) that is identical to
the model in Equation \ref{eq:m1b}, except that it excludes the logarithmic-transformed variable \textit{CT\_concepts} and the associated parameter $\beta_{11}$. The dependent variable is as follows:

\begin{dmath}
\log{\left(\frac{\hat{p}(\mathit{Stayed?})}{1-\hat{p}(\mathit{Stayed?})}\right)}
\end{dmath}




To answer R3 with the binary outcome variable \textit{Received_new_loves?}, we fit a logistic regression model on the full dataset. Our formal model (M3) is the same as that in Equation \ref{eq:m1b} but with the following dependent variable:

\begin{dmath}
\log{\left(\frac{\hat{p}(\mathit{Received\_new\_loves?})}{1-\hat{p}(\mathit{Received\_new\_loves?})}\right)}
\end{dmath}





\section{Results}
\label{sec:results}

\begin{figure*}[t]
    \centering
    \includegraphics[width=0.95\textwidth]{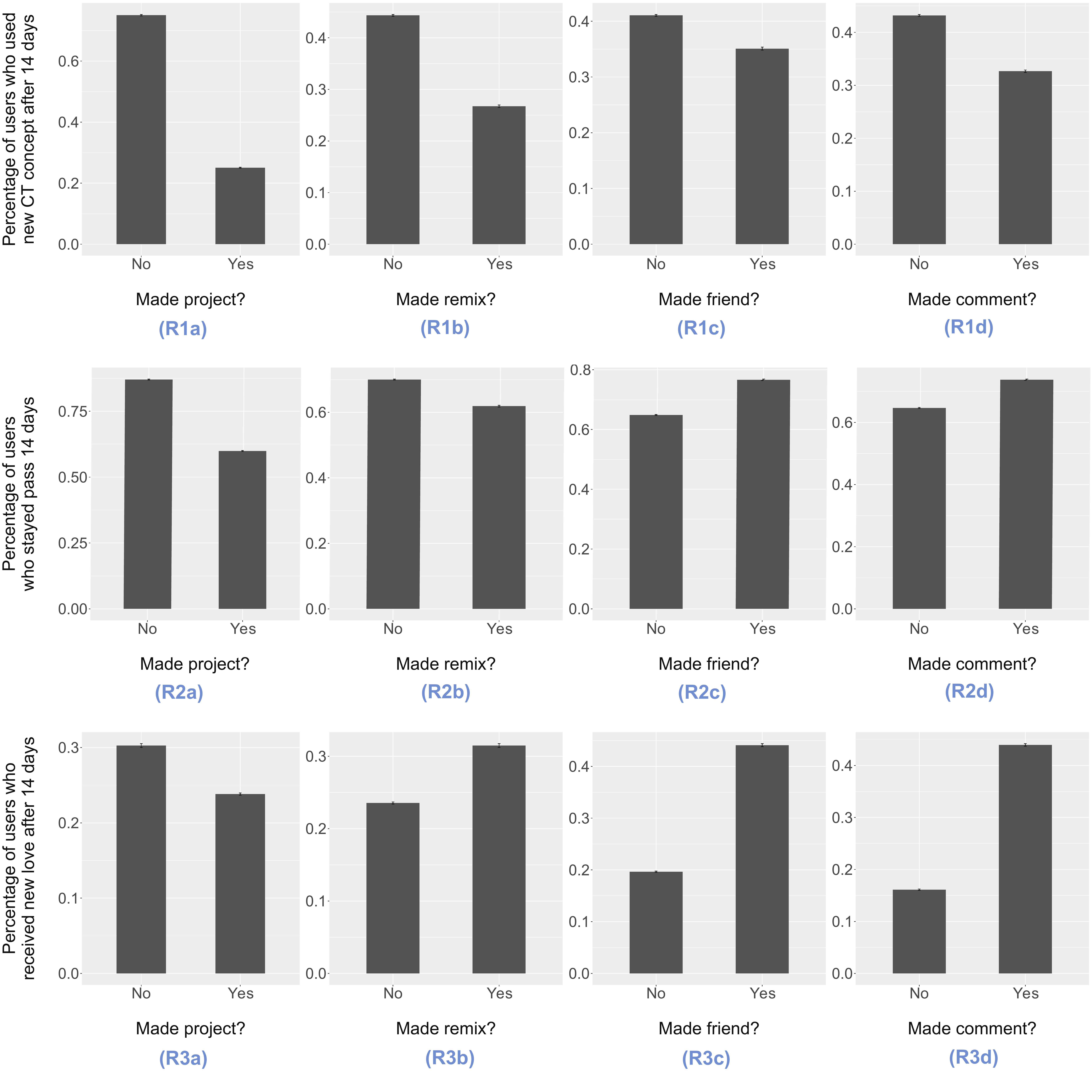}
    \caption{Bivariate plots that show the differences in the distributions of outcome variables across strata of our dataset that reflect the independent variables associated with each of our research questions.}
    \Description{There are three rows of bar plots in this graph. Each row is for a research question (marked as R1 to R3) and differs in y axis, (from top to bottom) Percentage of users who used new CT concept after 14 days, Percentage of users who stayed passed 14 days, and Percentage of users who received new love after 14 days. Each row contains four bar plots and is for an outcome variable (marked as a to d), which differ in x axis: (from left to right) Made project?, Made remix?, Made friend?, and Made comment?. The values of x axis are binary: No and Yes.
    
    The follows are information about values in each plot: 
    
    In R1a, the percentage of users who used new CT concepts after 14 days for users who did not make project in the first 14 days is around 0.8, while that for users who did is around 0.2. 
    
    In R1b, the percentage of users who used new CT concepts after 14 days for users who did not remix in the first 14 days is around 0.5, while that for users who did is around 0.25. 
    
     In R1c, the percentage of users who used new CT concepts after 14 days for users who did make friend in the first 14 days is around 0.4, while that for users who did is around 0.35. 
     
      In R1d, the percentage of users who used new CT concepts after 14 days for users who did not comment in the first 14 days is around 0.4, while that for users who did is around 0.3. 
      
       In R2a, the percentage of users who stayed past 14 days for users who did not make project in the first 14 days is around 0.9, while that for users who did is around 0.6. 
    
    In R2b, the percentage of users who stayed past 14 days for users who did not remix in the first 14 days is around 0.7, while that for users who did is around 0.6. 
    
    In R2c, the percentage of users who stayed past 14 days for users who did not make friends in the first 14 days is around 0.6, while that for users who did is around 0.8. 
    
    In R2d, the percentage of users who stayed past 14 days for users who did not comment in the first 14 days is around 0.6, while that for users who did is around 0.7. 
    
     In R3a, the percentage of users who received new loves after 14 days for users who did not make project in the first 14 days is around 0.3, while that for users who did is around 0.25. 
     
     In R3b, the percentage of users who received new loves after 14 days for users who did not remix in the first 14 days is around 0.2, while that for users who did is around 0.3. 
     
     In R3c, the percentage of users who received new loves after 14 days for users who did not made friends in the first 14 days is around 0.2, while that for users who did is around 0.4. 
     
     In R3d, the percentage of users who received new loves after 14 days for users who did not comment in the first 14 days is around 0.15, while that for users who did is around 0.4. 
    }
    \label{fig:eda_binary}
\end{figure*}


As the first step in our analysis, we construct a series of bivariate graphs to show differences in the distributions of our outcomes across strata representing our key independent variables. These plots are shown in Figure \ref{fig:eda_binary}.
In general, we find large differences in answers to all three of our research questions. 
Of course, these plots are exploratory; the observed relationships could be driven by any number of confounders. As a result, we explore our research questions using the full array of controls in our regressions. 

\begin{table}[t]
\begin{center}
\begin{tabular}{lccc}
\hline
 & \textit{New CT concept?} & \textit{Stayed?} & \textit{Received new love?} \\
& (M1) & (M2) & (M3) \\
\hline
(Intercept)                       & $1.47^{***}$ & $2.27^{***}$ & $-1.16^{***}$ \\
                                  & $(0.01)$  & $(0.02)$ & $(0.01)$     \\
\textit{Made\_original\_project?} & $0.44^{***}$ & $-0.63^{***}$ & $-0.71^{***}$  \\
                                  & $(0.03)$ & $(0.02)$ & $(0.04)$      \\
\textit{Made\_remix?}             & $-0.30^{***}$ & $-0.24^{***}$ & $0.00$  \\
                                  & $(0.02)$ & $(0.02)$ & $(0.02)$      \\
\textit{Made\_friend?}            & $0.01$ & $0.34^{***}$ & $0.28^{***}$  \\
                                  & $(0.02)$ & $(0.02)$ & $(0.02)$      \\                                
\textit{Made\_comment?}           & $0.00$ & $0.48^{***}$ & $0.95^{***}$  \\
                                  & $(0.02)$ & $(0.02)$ & $(0.02)$      \\
\textit{Has\_been\_friended?}         & $0.20^{***}$ & $0.38^{***}$ & $0.42^{***}$  \\
                                  & $(0.02)$ & $(0.02)$ & $(0.02)$      \\
\textit{Has\_been\_loved?}            & $-0.02$ & $0.19^{***}$ & $0.70^{***}$  \\
                                  & $(0.03)$ & $(0.02)$ & $(0.02)$      \\
\textit{Has\_been\_viewed?}           & $-1.25^{***}$ & $-1.70^{***}$ & $-0.76^{***}$      \\
                                  & $(0.02)$ & $(0.03)$ & $(0.03)$      \\
\textit{Has\_been\_remixed?}          & $0.04$ & $0.12^{***}$ & $0.31^{***}$  \\
                                  & $(0.03)$ & $(0.03)$ & $(0.03)$      \\
\textit{Has\_been\_commented?}        & $-0.13^{***}$ & $0.00$ & $0.26^{***}$  \\
                                  & $(0.02)$ & $(0.02)$ & $(0.02)$     \\
\textit{Has\_been\_favorited?}        & $-0.04$ & $0.17^{***}$ & $0.41^{***}$  \\
                                  & $(0.03)$ & $(0.03)$ & $(0.02)$      \\
log1p(\textit{CT\_concepts})  & $-1.19^{***}$ & N/A & $0.17^{***}$ \\
                                  & $(0.02)$ &  & $(0.02)$    \\                        
\hline
AIC                               & $125313.82$ & $133601.48$ & $119286.65$  \\
BIC                               & $125430.28$ & $133708.23$ & $119403.11$  \\
Log Likelihood                    & $-62644.91$ & $-66789.74$ & $-59631.33$  \\
Deviance                          & $125289.82$ & $133579.48$ & $119262.65$  \\
McFadden's pseudo R-squared       & $0.23$ & $0.12$ & $0.14$  \\
Num. obs.                         & $121149$ & $121149$ & $121149$    \\
\hline
\multicolumn{2}{l}{\scriptsize{$^{***}p<0.001$; $^{**}p<0.01$}; $^{*}p<0.05$}
\end{tabular}
\caption{Logistic regression models that predict the probability that a user uses a new CT concept (M1), stays active (M2), and receives loves on new projects (M3) created after their first 14 days on the community. The models are fit to the user-level dataset that includes aggregated activities of $121,149$ users.}
\label{table:results_binary}
\end{center}
\end{table}

\begin{figure*}[t]
    \centering
    \includegraphics[width=0.95\textwidth]{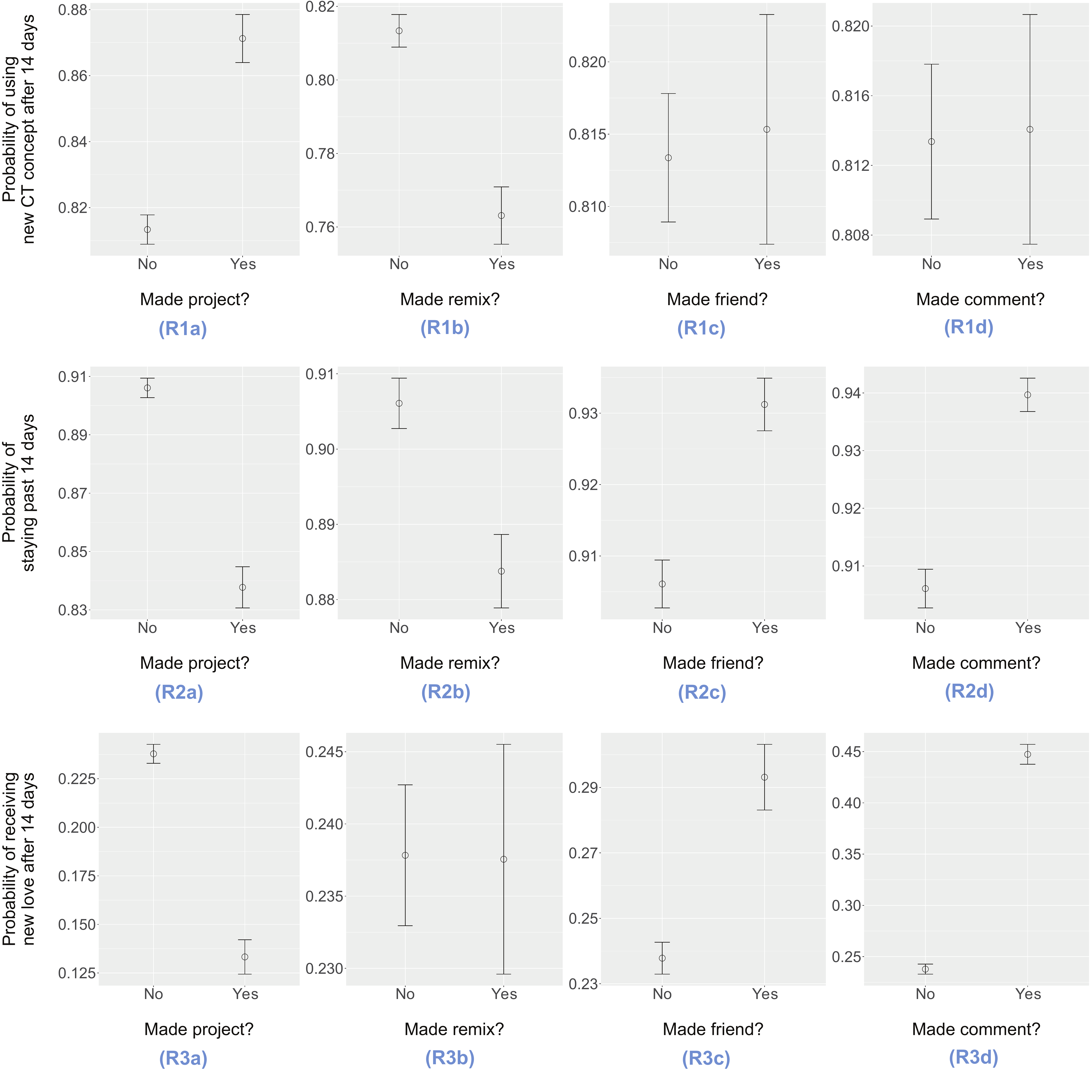}
    \caption{Plots of model predicted probabilities corresponding to each of our research questions. Each figure shows the model predicted probabilities for two prototypical users who have median values for each of our control variables; these users vary only in terms of the key independent variable in the corresponding hypothesis. Error bars reflect the marginal effects of our key independent variable ($1.96\times\mathrm{SE}$).}
    \Description{There are three rows of plots in this graph. Each row is for a research question (marked as R1 to R3) and differs in their $y$ axis: (from top to bottom) Probability of using new CT concept after 14 days, Probability of staying passed 14 days, and Probability of receiving new love after 14 days.
    
    Each row contains four plots and is for an outcome variable (marked as a to d) which differ in $x$ axis: (from left to right) Made project?, Made remix?, Made friend?, and Made comment?. The values of x axis are for the two category of prototypical users: users who have not done the corresponding activity in their first 14 days (marked as No) and users who have  done the corresponding activity in their first 14 days (marked as Yes).
    
    Inside each plot are two dots representing model prediction value with error bar for both the prototypical users. The specific values in each plot are described in the text in the result section.}
    \label{fig:regression_binary}
\end{figure*} 

The regression results for all three research questions are shown in Table \ref{table:results_binary}.
Because interpreting marginal effects from regression models with many covariates can be challenging, especially in the case of GLM models such as logistic regression, we present a series of plots of model-predicted values in Figure \ref{fig:regression_binary}. We do so by identifying the median values for each of our control variables (shown in Table \ref{table:distribution_variables}) and then generating model predicted values for two prototypical users that vary only in terms of our key independent variables---e.g., a user who has (or has not) remixed, and so on. Each of the key independent variables not being visualized is also held at the sample medians.

\subsection{Domain}

For R1, we found that making original projects as a newcomer is positively associated with the usage of new computational concepts in the long term, whereas remixing as a newcomer has the opposite effect. 
As shown in Table \ref{table:results_binary}, the odds that a user who created an original project in their first 14 days on Scratch would use new CT concepts afterward are 1.55 times the odds that an otherwise similar user who had not created an original project would do so ($\beta=0.44$; $\mathrm{SE}=0.03$; $p<0.001$). As seen in Figure \ref{fig:regression_binary} (R1a), our model predicts that approximately 81\% of the prototypical users who had not made original projects in their first 14 days would use new CT concepts afterward, whereas approximately 87\% of otherwise similar users who had made original projects would. Furthermore, we find that the odds that a user who remixed in their first 14 days would use a new CT concept afterward are 0.74 times the odds of an otherwise similar user who had not done so ($\beta=-0.30$; $\mathrm{SE}=0.02$; $p<0.001$). As seen in Figure \ref{fig:regression_binary} (R1b), our model predicts that approximately 81\% of prototypical users who had not remixed in their first 14 days would use new CT concepts afterward, whereas approximately 76\% of the otherwise similar users who had remixed would. The effects of friending and commenting are not statistically significant. 


\subsection{Community}

For R2, we found that making original projects and remixing as a newcomer are negatively associated with long-term stay in the community, whereas making friends and commenting as a newcomer were positively correlated with the probability of staying in the community for 14 days. 
The odds that a user who created an original project in their first 14 days in Scratch would stay in the community afterward are 0.53 times the odds that an otherwise similar user who had not created an original project would do so ($\beta=-0.63$; $\mathrm{SE}=0.02$; $p<0.001$). As seen in Figure \ref{fig:regression_binary} (R2a), our model predicts that approximately 91\% of prototypical users who had not made original projects in the first 14 days will stay longer than 14 days, whereas only approximately 84\% of otherwise similar users who had made original projects would stay. In terms of the effect of remixing, the odds that users who remixed others' projects in their first 14 days will stay longer than 14 days are 0.79 times the odds of a user who did not do so ($\beta=-0.24$; $\mathrm{SE}=0.01$; $p<0.001$). As seen in Figure \ref{fig:regression_binary} (R2b), our model predicts that more than 90\% of prototypical users who did not remix in their first 14 days would stay in the community beyond 14 days, whereas approximately 88\% of otherwise similar users who had remixed would. 
In contrast, the odds that users who friended others in their first 14 days would stay past that time period are 1.40 times the odds that a user who did not friend anyone would ($\beta=0.34$; $\mathrm{SE}=0.02$; $p<0.001$). Figure \ref{fig:regression_binary} (R2c) shows that our model predicts that approximately 90\% of our prototypical users who had not friended anyone in their first 14 days would stay in the community past 14 days, whereas approximately 93\% of otherwise similar users who had friended people would do so. For users who posted feedback in the community in their first 14 days, their odds of staying in Scratch after 14 days are 1.62 times the odds of users who did not ($\beta=0.48$; $\mathrm{SE}=0.01$; $p<0.001$). Figure \ref{fig:regression_binary} (R2d) shows that our model predicts that only approximately 90\% of prototypical users who had not posted any comments in their first 14 days would stay in the community past 14 days, whereas approximately 94\% of otherwise similar users who had posted comments would.

\subsection{Practice}
For R3, we found that while making original projects as a newcomer is negatively associated with receiving loves from the community in the long term, making friends and commenting as a newcomer have a positive relationship with the outcome.
Our model predicts that users who created an original project in their first 14 days in Scratch have 0.49 times the odds of receiving loves for their projects after 14 days compared with those who did not post feedback in the same newcomer period ($\beta=-0.71$; $\mathrm{SE}=0.04$; $p<0.001$). As seen in Figure \ref{fig:regression_binary} (R3a), our model predicts that approximately 24\% of prototypical users who had not made original projects in their first 14 days will stay longer than 14 days, whereas only approximately 13\% of otherwise similar users who had made original projects would.
In contrast, the odds that users who friended others in their first 14 days would receive new loves are 1.32 times the odds that a user who did not friend anyone would do so ($\beta=0.28$; $\mathrm{SE}=0.02$; $p<0.001$). Figure \ref{fig:regression_binary} (R3c) shows that our model predicts that approximately 24\% of our prototypical users who had not friended anyone in the first 14 days would stay in the community past 14 days, whereas approximately 29\% of otherwise similar users who had friended people would do so.
In terms of feedback exchange, we find evidence that users who posted feedback in the community in their first 14 days would have 2.58 times the odds to receive loves for their projects after 14 days compared to those who did not post feedback in their newcomer period ($\beta=0.95$; $\mathrm{SE}=0.02$; $p<0.001$). As shown in Figure \ref{fig:regression_binary} (R3d), the predicted probability of receiving loves for new projects created after 14 days is only approximately 25\% for prototypical users who did not comment during their initial period, but this value is close to 45\% for those who did. The effect of remixing is not statistically significant.


\section{Discussion}
\label{sec:discussion}

In this paper, we present a quantitative study with data from the Scratch community that investigates how newcomers' LPP contributes to different kinds of learning in CoP. Our paper makes a contribution to social computing theory by drawing in \citepos{wenger2002cultivating} three types of learning from the broader CoP literature: development of domain skills, development of identity as a community member, and development of community-specific values and practices. In our synthesis of the CoP literature in social computing, we identified four types of LPP common among newcomers (contribution to core tasks, engagement with practice proxies, social bonding, and feedback exchange) and formed three research questions related to understanding how the early participation of users in the community will predict learning outcomes in CoPs.

We contribute what we think is the first quantitative test of CoP theory in the context of social computing. 
Analytically speaking, our work tested 12 hypotheses that correspond to the 12 possible relationships between our independent and dependent variables (i.e., the lines in Figure \ref{fig:research questions}). Our work finds that most of these relationships are statistically significant, but that the signs and relative magnitudes of parameters associated with each type of LPP vary across the types of learning outcomes. At a very high level, our results provide concrete evidence that different types of LPP have different relationships with different important facets of learning in online CoPs. 
What is productive for some types of learning outcomes is unhelpful for others, and vice versa.
We devote the remainder of our discussion to unpacking our specific results.

\subsection{Supporting contribution to core tasks}

Our study finds quantitative evidence for the belief that learners in a CoP will learn the domain by being part of the core tasks in the community. This finding is consistent with the constructionist design approach behind Scratch. In appears that by participating in epistemologically relevant project creation, learners can appropriate the abstract concepts behind their hands-on experiences \cite{papert1980mindstorms, resnick1996pianos, resnick2009scratch}. 

That being said, contribution to core tasks is negatively associated with our measure of learning as membership identity and community-specific practices. We speculate that because creating a programming project from scratch can be challenging for novices, some newcomers might feel discouraged and, as a result, turn away from the community. This is consistent with the results reported in previous literature, which indicate that it can be challenging for newcomers to follow community norms while making contributions \cite{halfaker2013rise}. Existing resources reflecting good practices are often composed by experienced users and contain contextual information that is not beginner-friendly \cite{marlow2014rookie, fiesler2017growing}. These might be obstacles for newcomers looking to adopt community values. 

Designers of the Scratch language have made enormous efforts to lower the bar of programming for novices \cite{resnick2009scratch}. Still, making a new project, however simple, might be overwhelming for someone without programming experience. Future design of informal learning communities like Scratch could consider offering easier options for newcomers to engage in core tasks. For example, in some other online CoPs dedicated to software development, newcomers can participate in simple, collaborative, programming-related tasks such as reporting bugs \cite{Krogha2003CommunityJA}.
Scratch might try to direct newcomers toward types of newcomer-friendly programming tasks such as building on example templates, identifying bugs in collaborative projects, and helping more experienced users with a small part of their projects.

\subsection{Supporting engagement with practice proxies}

Our exploration shows that engagement with practice proxies is negatively associated with the development of domain knowledge and the formation of membership identity. Although previous work on Scratch shows that users who remixed projects containing a CT concept tended to use the same CT concept in their own projects \cite{dasgupta2016remixing}, our findings suggest that this may not contribute to the expansion of computational knowledge and skills in general. Our results are consistent with previous findings on remixing, in which users tend to show less innovation and originality when building from the most frequently remixed programs \cite{hill2013cost}. 


More research needs to be conducted to understand exactly why remixing negatively impacts some dimensions of learning. One possible explanation is that newcomers may have trouble understanding specific choices in projects made by more experienced members and may not have the confidence needed to try and build upon them. In Kaggle, novices face a similar challenge of understanding and reusing publicly shared code by experts due to incomplete documentation on dependencies, missing rationales, and lack of context \cite{cheng2020building}. In these cases, simply providing newcomers with practice proxies is likely not enough.

Possible solutions might include additional scaffolding by established community members to document their steps and reasoning processes in the first place. \citet{wang2020callisto} and \citet{rule2018aiding} provide great guidelines and examples for such design in the context of computational learning. Informal learning communities should also offer newcomers the opportunity to ask questions to experienced members. Such a design should scaffold the asker to clearly communicate which step they are confused about in context and support answerers to effectively showcase their procedures and rationales. For example, in the context of Scratch, a synchronous Q\&A function (e.g., an opt-in chat room similar to the one proposed by \citet{ford2018we}) could be implemented in the remix feature so that newcomers could discuss the coding mechanisms with the original creators of the projects being remixed. 

\subsection{Supporting newcomer's socialization}
Newcomers' social bonding activity is associated with the formation of community identity. Such socialization activities are also positively correlated with the development of community-specific practices. This finding can be read as a strong response to the skeptics of informal learning in online CoPs who argue that chatting and socialization will distract users. Our results complement other work that shows that socialization can be a legitimate pathway to learning what the community values \cite{cheng2022how}. One possible interpretation is that social bonding can facilitate computational participation \cite{kafai_computational_2016}---learning through participating in socially situated contexts. 

Better ways to socialize newcomers is a perennial topic in social computing.
Our results suggest that success in these efforts can support a variety of learning outcomes.
Proactive efforts to reach out to new users with comments are a useful way to jumpstart social bonding processes.
Designers of informal online learning systems may consider intentionally helping newcomers socialize with each other and older members through community events or algorithmic matching mechanisms, for example, making the transition from identity-based connections to bond-based connections possible \cite{kraut_building_2011}.   

\subsection{Supporting feedback exchange}

Feedback exchange is positively associated with our measures of learning about both community and practice. Complementing previous studies of collaborative debugging \cite{shorey2020hanging}, our study offers quantitative evidence that socially oriented newcomer participation can contribute to community practices. 
Despite the promise of feedback exchange on learning, newcomers refrain from publicly offering feedback because they are self-conscious about their social status and are not confident in their expertise \cite{marlow2014rookie}. Nonpublic or anonymous feedback systems might be established so that newcomers do not have to reveal themselves while interacting with community members \cite{ford2018we}. Alternatively, scaffolds for composing discussion messages could be introduced so that newcomers can have some idea about what to say in comments and participate in feedback exchange more easily \cite{hui2018introassist, macneil_framing_2021}.




\section{Threats to Validity}
\label{sec:threats}

A series of articles in the social computing literature have argued about whether deeply engaged and highly active users in online communities are ``born and not made'' \citep{huang2015activists, panciera2009wikipedians, preece2009reader}. In its analytic structure, our work has quite a bit in common with \textit{born versus made} studies by \citet{panciera2009wikipedians} and \citet{huang2015activists}. Both studies seek to make predictions about users' long-term behavior using measures of early activity. There are also important differences between these approaches and ours. Born/made studies typically look at users' very first engagement. Although we take a longer view in the construction of our measures, our results are not sensitive to this decision.
Although we include a large body of control variables to address \textit{ex ante} differences between users, we cannot fully rule out the explanation that users who are more active early on are systematically different from users who are not.
In contrast, we might argue that our work points to an alternative explanation for early studies that suggest that users are ``born and not made.''
Both our theoretical framework and empirical results suggest that early activity may not be best thought of as an indicator of existing differences in knowledge, commitment, and skills, but rather as a pathway through which users can develop. 

Furthermore, we define the ``newcomer'' period as the first 14 days after a user creates an account. Although this decision is informed by our experiences with Scratch and although robustness checks on user activities in the first 2, 7, and 30 days lead to consistent results, the choice of 14 days is arbitrary and fails to capture important nuance in the way that Scratch users understand their own status, and the status of others, within their community. We hope that future work can improve on this very limited measurement approach.

Another limitation of our work is that the four types of LPP that we identified are not a comprehensive list of all possible types of LPP in online communities, either in general or in Scratch in particular. As the result of our review of relevant social computing literature, our enumeration of these four types is put forward with humility and knowledge that is limited and incomplete. Although we know that our list is not comprehensive, we hope that our enumeration and measurement of four distinct types of LPP are sufficient to convey our high-level argument that there are different types of newcomer engagement that can lead to different learning pathways. Future research could explore other types of LPP and their impact on long-term learning outcomes.

Our work is also limited in that our measures are proxies rather than direct measures of the concepts of interest.
This is particularly important to acknowledge in terms of our learning outcome measures. We rely on proxies because we simply do not have a way to evaluate actual learning in Scratch.
For example, we use differences in incoming loves as a proxy for an individual's learning about community-specific practices and values. It is important to remember that receiving positive community reactions is an outcome of learning community practices, not a direct measure of learning. Another example is that we use a measure of whether a user stays in the community for longer than 14 days to proxy for a learner's development of membership identity. We rely on proxies because sociopsychological concepts are difficult to directly measure in informal settings like Scratch, where there is little or no ability to conduct tests or even communicate with users. Although we draw heavily from other work that has developed and evaluated proxy measures in Scratch, all of these attempts are risky and prone to noise, or worse \citep{dasgupta2016remixing, dasgupta2018wide}. Although far from perfect, we believe that these proxies provide valuable insight into our research questions.

While our research questions describe causal relations, our regression analyses can only provide correlational evidence. As is often the case in cross-sectional analyses, these relationships might be due to variables that are correlated with, but not caused by, our predictors. We hope that design-based field experiments can be conducted in the future to identify causal relationships between specific types of newcomer LPP and the specific kind of learning in CoPs. Until then, we offer our results as tentative evidence. Furthermore, our analysis focuses only on the main effect of the independent variables. In an earlier version of the analysis, we include two-way interactions of our independent variables as predictors in our models, and we include the results in Table \ref{table:interaction} in the appendix. Since understanding the interaction relationships between our independent variables is not the goal of this analysis and it is difficult to interpret the coefficients of interactions, we did not integrate these models into our main analysis. We urge future researchers to build on our work and explore the interactions and combination of different types of LPP and their contribution to different types of learning.


Our work is also limited in that all our outcomes are dichotomous. This is a choice we made for two reasons. First, it means that our analysis is at a lower risk of violating parametric modeling assumptions. Second, doing so also captures the vast majority of the variation in our outcomes. However, this choice means that our results only paint part of the picture. Although we show that sharing new projects and remixing early on are associated with a higher chance of using CT concepts later, we do not know if more remixing leads to more new CT concepts among users who continue.
Although we leave this question for future work, we include an exploratory analysis in Table \ref{table:results_continuous} in our appendix that reproduces our three models with continuous measures of our dependent variables among the relatively small proportion of users for whom our outcomes were equal to 1. This is conceptually similar to a hurdle model.
Although limited in several ways, these results suggest that our dichotomous measures, which are strong predictors of new CT concepts, staying active, and receiving loves, are not necessarily strong predictors of higher amounts of learning among users who have learned at all.
We hope to refine and continue to explore the relationship between the types of newcomer LPP and the magnitude of learning outcomes in future studies.

Finally, our analysis focuses on Scratch as the single empirical setting. While Scratch is large, active, and supports a wide range of activities, it is also unique in many aspects: its user population mostly consists of young people, the community focuses on computational learning, and it contains affordances such as
remixing and viewing others' full code. We do not know if our findings from the Scratch online community will generalize to other settings or population groups. In addition, we analyzed activities during a single time period, which was early in the Scratch community's lifetime. Since the affordances of Scratch have changed and the community itself has also grown since then, we cannot know how well our findings will translate to Scratch today. We hope that our case study offers a framework for studying different types of LPP and learning in online CoPs. Future researchers could answer our research questions in other settings and compare their results with our findings. 


\section{Conclusion}
\label{sec:conclusion}
In this study, we present a quantitative approach to engaging with CoP theory in an informal online learning context. Although previous studies in social computing have treated learning outcomes as a single dimension, we build on work by \citet{wenger2002cultivating} to describe three distinct types of learning that can be supported in online CoPs: learning about the domain, learning about the community, and learning about the practice. We also identify four forms of LPP that are common in informal online learning contexts. We use historical data from Scratch to answer a series of research questions about whether certain types of newcomer LPP contribute to certain kinds of learning. Our results suggest that there is a range of possible pathways to a range of distinct learning outcomes.

Taking a broad exploratory approach instead of a more narrow hypothesis testing structure, we find evidence of relationships between types of LPP and learning outcomes that we believe are entirely untheorized in the CoP literature in social computing. For example, we find that newcomer socializing is associated with learning about the community and its practices. Our study offers theoretical and empirical contributions to social computing research on informal learning settings, as well as practical implications on how to best design informal online learning systems. In summary, our study suggests that online communities afford many possible destinations in learning and many pathways to each.  


\newpage

\begin{acks}
We would like to acknowledge members of the Community Data Science Collective for their feedback and support. We also want to thank our anonymous reviewers for their feedback. We want to thank the Scratch team---especially Mitchel Resnick,  Natalie Rusk, and Andrés Monroy-Hernandez---for building the Scratch online communities and for making data about Scratch available that allowed us to conduct this work. Finally, we want to thank members of the Scratch online community who generated data for this research and who inspired our work.
\end{acks}

\bibliographystyle{ACM-Reference-Format}
\bibliography{sample-base}

\newpage
\appendix

\section{Appendix}

\begin{table}[h]
\begin{center}
\begin{tabular}{l l }
\hline
CT concepts & Scratch Blocks \\
\hline
Loops & forever, foreverIf, repeat, repeatUntil \\
Parallelism & startHatTriggered, eventHatTriggered, keyHatTriggered, mouseHatTriggered \\
Events & eventHatTriggered, keyHatTriggered, mouseHatTriggered, bounceOffEdge, \\ 
& turnAwayFromEdge, touching, touchingColor, colorSees, mousePressed, \\ 
& keyPressed, isLoud, sensor, sensorPressed, distanceTo \\
Conditionals & waitUntil, foreverIf, if, ifElse, repeatUntil, bounceOffEdge, \\
 & turnAwayFromEdge, touching, touchingColor, colorSees, mousePressed, \\
 & keyPressed, isLoud, sensor, sensorPressed, lessThan, equalTo, greaterThan, \\
 & and, or, not, listContains \\
Operators & lessThan, equalTo, greaterThan, and, or, not, add, \\
& subtract, multiply, divide, pickRandomFromTo, concatenateWith,\\
& letterOf, stringLength, mod, round, abs, sqrt, \\
& sin, cos, tan, asin, acos, atan, ln, log, eˆ, 10ˆ\\

Data & setVarTo, changeVarBy, showVariable, hideVariable, readVariable, \\
 & addToList, deleteLineOfList, insertAtOfList, setLineOfListTo,\\ & contentsOfList, getLineOfList, lineCountOfList, listContains \\
\hline
\end{tabular}
\caption{CT concepts mapping to blocks in the Scratch programming language, adopted from \citet{dasgupta2016remixing}. We use this mapping to construct our outcome variable for H1a and H1b about whether new CT concept is presented in projects created after a user's first 14 days in the community.}
\label{table:CT}
\end{center}
\end{table}

\begin{table}[h]
\begin{center}
\begin{tabular}{l c c c}
\hline
Variables & Total new CT concepts & Active duration & New loves per project \\
& (M1) & (M2) & (M3) \\
\hline
(Intercept)                       & $4.65^{**}$ & $4.81^{**}$ & $0.15^{**}$  \\
                                  & $(0.01)$ & $(0.01)$ & $(0.00)$      \\
Made\_original\_project? & $-0.09^{*}$ & $-0.28^{**}$ & $-0.15^{**}$ \\
                                  & $(0.04)$ & $(0.02)$ & $(0.01)$      \\                                   
Made\_remix?             & $0.02$ & $-0.04^{*}$ & $-0.04^{**}$  \\
                                  & $(0.02)$ & $(0.01)$ & $(0.00)$      \\
Made\_friend?            & $-0.18^{**}$ & $-0.03^{*}$ & $0.02^{**}$  \\
                                  & $(0.02)$ & $(0.01)$ & $(0.00)$      \\
Made\_comment?           & $-0.09^{**}$ & $0.32^{**}$ & $0.16^{**}$  \\
                                  & $(0.02)$ & $(0.01)$ & $(0.00)$     \\
Has\_been\_friended?         & $-0.06^{*}$ & $0.09^{**}$ & $0.08^{**}$  \\
                                  & $(0.02)$ & $(0.01)$ & $(0.00)$     \\
Has\_been\_loved?            & $-0.03$ & $0.11^{**}$ & $0.19^{**}$  \\
                                  & $(0.03)$ & $(0.01)$ & $(0.01)$      \\
Has\_been\_viewed?           & $0.00$ & $0.06^{**}$ & $0.00$  \\
                                  & $(0.03)$ & $(0.02)$ & $(0.01)$      \\
Has\_been\_remixed?          & $-0.02$ & $0.11^{**}$ & $0.05^{**}$  \\
                                  & $(0.03)$ & $(0.02)$ & $(0.01)$      \\
Has\_been\_commented?        & $0.01$ & $0.14^{**}$ & $0.01^{*}$  \\
                                  & $(0.02)$ & $(0.01)$ & $(0.00)$      \\
Has\_been\_favorited?        & $0.01$ & $0.11^{**}$ & $0.12^{**}$  \\
                                  & $(0.03)$ & $(0.02)$ & $(0.01)$      \\
log1p(CT\_concepts)  & $-1.68^{**}$ & NA & $0.03^{**}$\\
                                  & $(0.02)$ & NA & $(0.00)$     \\
\hline
R$^2$                             & $0.47$ & NA & $0.15$      \\
Adj. R$^2$                        & $0.47$ & NA & $0.15$      \\
AIC                               & NA & $1140546.37$ & NA \\
BIC                               & NA & $1140660.20$ & NA \\
Log Likelihood                    & NA & $-570261.19$ & NA \\
Deviance                          & NA & $116442.92$ & NA  \\
McFadden's pseudo R-squared       & $0.16$ & $0.01$ & $0.14$  \\
Num. obs.                         & $47952$ & $97295$ & $79963$      \\
\hline
\multicolumn{2}{l}{\scriptsize{$^{***}p<0.001$; $^{**}p<0.01$}; $^{*}p<0.05$}
\end{tabular}
\caption{Results from our early exploratory analysis based on continuous construct of outcome variables and subsets of our user level dataset used from the main analysis. M1_c tests H1, where \textit{Total_new_CT_concepts} is a count variable of number of new CT concepts used in projects created after the first 14 days among the $47,952$ users who had used new CT concepts after 14 days. M2_c tests H2, where \textit{Active_duration} is a count variable of the number of days that a user engages in any recorded community activities and is measured by the number of days between the the end of the users 14 day period and the last in which they are active. This is also constructed on the subset of $97,295$ users who were active after the first 14 days. M3_c tests H3, where \textit{New_loves_per_project} is a continuous variable of average number of loves received by a user on projects created after the first 14 days, among the $79,963$ users who did create projects after 14 days. We chose to exclude these results from our main findings due to small sample size (< 10\% of our original dataset), and truncation issues.}
\label{table:results_continuous}
\end{center}
\end{table}

\begin{table}
\begin{center}
\begin{tabular}{l c c c}
\hline
    Variables & New CT concept?  & Stayed? & Received new love? \\
& (M1) & (M2) & (M3) \\
\hline
(Intercept)                                                 & $1.83^{***}$ & $3.39^{***}$ & $-1.04^{***}$\\
                                                          & $(0.02)$ & $(0.04)$ & $(0.02)$     \\
Made\_original\_project?                         & $-0.79^{***}$ & $-3.05^{***}$ & $-1.63^{***}$ \\
                                                          & $(0.04)$ & $(0.05)$ & $(0.06)$     \\                           
Made\_remix?                                     & $-2.34^{***}$ & $-3.59^{***}$ & $-1.54^{***}$ \\
                                                          & $(0.04)$  & $(0.05)$ & $(0.06)$    \\
Made\_friend?                                    & $-0.08$ & $0.30^{***}$ & $0.21^{***}$ \\
                                                          & $(0.05)$ & $(0.07)$  & $(0.04)$   \\ 
Made\_comment?                                    & $0.03$ & $0.52^{***}$ & $1.01^{***}$ \\
                                                          & $(0.04)$ & $(0.06)$ & $(0.03)$     \\
Has\_been\_friended?                                 & $0.17^{***}$ & $0.35^{***}$ & $0.40^{***}$ \\
                                                          & $(0.02)$ & $(0.02)$ & $(0.02)$     \\
Has\_been\_loved?                                    & $-0.05$ & $0.16^{***}$ & $0.68^{***}$ \\
                                                          & $(0.03)$ & $(0.02)$ & $(0.02)$    \\
Has\_been\_viewed?                                    & $-0.27^{***}$ & $-0.35^{***}$ & $0.14^{**}$\\
                                                           & $(0.03)$ & $(0.03)$ & $(0.05)$     \\
Has\_been\_remixed?                                   & $-0.01$ & $0.08^{*}$ & $0.29^{***}$  \\
                                                           & $(0.03)$ & $(0.03)$ & $(0.03)$   \\
Has\_been\_commented?                                 & $-0.13^{***}$ & $-0.01$ & $0.25^{***}$     \\
                                                           & $(0.02)$ & $(0.02)$ & $(0.02)$    \\
Has\_been\_favoriated?                                & $-0.09^{**}$ & $0.13^{***}$ & $0.39^{***}$ \\
                                                          & $(0.03)$ & $(0.03)$ & $(0.02)$    \\
log1p(CT\_concepts)                          & $-1.22^{***}$ & NA & $0.13^{***}$\\
                                                          & $(0.02)$ & NA & $(0.02)$     \\                                                          
Made\_comment? × Made\_remix?             & $0.22^{***}$ & $0.17^{***}$ & $0.11^{**}$\\
                                                          & $(0.04)$ & $(0.04)$ & $(0.04)$    \\
Made\_comment? × Made\_friend?             & $-0.01$ & $0.11^{**}$ & $-0.12^{***}$ \\
                                                          & $(0.04)$ & $(0.04)$ & $(0.04)$    \\
Made\_comment? × Made\_original\_project?  & $-0.20^{***}$ & $-0.19^{***}$ & $-0.10^{**}$\\
                                                          & $(0.04)$ & $(0.06)$ & $(0.04)$    \\
Made\_remix? × Made\_friend?               & $0.10^{*}$ & $0.05$  & $0.15^{***}$     \\
                                                           & $(0.04)$ & $(0.04)$ & $(0.04)$     \\
Made\_remix? × Made\_original\_project?    & $2.30^{***}$ & $3.60^{***}$ & $1.59^{***}$\\
                                                           & $(0.05)$ & $(0.05)$ & $(0.06)$    \\
Made\_friend? × Made\_original\_project?  & $0.06$ & $-0.06$  & $0.13^{***}$     \\
                                                           & $(0.04)$ & $(0.06)$ & $(0.04)$    \\
\hline
AIC                                                       & $122375.61$ & $128298.77$ & $118346.10$  \\
BIC                                                       & $122550.30$ & $128463.75$  & $118520.78$ \\
Log Likelihood                                            & $-61169.80$ & $-64132.39$ & $-59155.05$  \\
Deviance                                                   & $122339.61$ & $128264.77$ & $118310.10$  \\
McFadden's pseudo R-squared       & $0.25$ & $0.16$ & $0.14$  \\
Num. obs.                                                 & $121149$ & $121149$ & $121149$     \\
\hline
\multicolumn{2}{l}{\scriptsize{$^{***}p<0.001$; $^{**}p<0.01$; $^{*}p<0.05$}}
\end{tabular}
\caption{Logistic regression models that include two-way interaction between independent variables as predictors. Same as our main analysis, the models predict the likelihood that a user using new CT concept (M1), staying active (M2), and receiving loves on new projects (M3) created after their first 14 days in the community. Models are fit on the user level dataset including aggregated activities from 121149 users.}
\label{table:interaction}
\end{center}
\end{table}

\end{document}